\documentclass[Referee]{aa}
\usepackage{graphicx}
\usepackage[varg]{txfonts}
\usepackage{cases}
\usepackage{natbib}
\graphicspath{{./}{pics/}}

\begin{document}
      \title{Overexpansion-dominated Coronal Mass Ejection Formation and Induced Radio Bursts}
      
      \author{B. T. Wang\inst{1,2}, X. Cheng\inst{1,2,3}, H. Q. Song\inst{4}, M. D. Ding\inst{1,2}}
      
      \institute{School of Astronomy and Space Science, Nanjing University, Nanjing 210023, People's Republic of China 
      	\email{xincheng@nju.edu.cn}
 		\and
 		Key Laboratory of Modern Astronomy and Astrophysics (Nanjing University), Ministry of Education, Nanjing 210093, People's Republic of China 
 		\and
 		Max Planck Institute for Solar System Research, Gottingen, 37077, Germany 
		\and
		Shandong Provincial Key Laboratory of Optical Astronomy and Solar-Terrestrial Environment, and Institute of Space Sciences, Shandong University, Weihai, Shandong 264209, People’s Republic of China
 	    }
    \titlerunning{Overexpansion of a Coronal Mass Ejection}
 \authorrunning{Wang et al.}
\date{Received; accepted}

\abstract
{}
{Coronal Mass Ejections (CMEs) are the most fascinating explosion in the solar system; however, their formation is still not fully understood.} {Here, we investigate a well-observed CME on 2021 May 07 that showed a typical three-component structure and was continuously observed from 0 to 3 R$_{\sun}$ by a combination of SDO/AIA (0--1.3 R$_{\sun}$), PROBA2/SWAP (0--1.7 R$_{\sun}$) and MLSO/K-Cor (1.05--3 R$_{\sun}$). Furthermore, we compare the morphological discrepancy between the CME white-light bright core and EUV blob. In the end, we explore the origin of various radio bursts closely related to the interaction of the CME overexpansion with nearby streamer.}
{An interesting finding is that the height increases of both the CME leading front and bright core are dominated by the overexpansion during the CME formation. The aspect ratios of the CME bubble and bright core, quantifying the overexpansion, are found to decrease as the SO/STIX 4--10 keV and GOES 1--8 {\AA} soft X-ray flux of the associated flare increases near the peaks, indicating an important role of the flare reconnection in the first overexpansion.
The CME bubble even takes place a second overexpansion although relatively weak, which is closely related to the compression with a nearby streamer and likely arises from an ideal MHD process.
Moreover, the CME EUV blob is found to be relatively lower and wider than the CME white-light bright core, may correspond to the bottom part of the growing CME flux rope. 
The interaction between the CME and the streamer leads to two type II radio bursts, one normally drifting and one stationary, which are speculated to be induced at two different sources of the CME-driven shock front. The bidirectional electrons evidenced by series of "C-shaped" type III bursts suggest that the interchange reconnection be also involved during the interaction of the CME and streamer.}
{}


\maketitle

\section{Introduction} \label{sec:intro}

Coronal Mass Ejections (CMEs) are the most fascinating solar activities occurring in the solar corona and are able to release a large amount of magnetized plasma to the interplanetary space. If propagating towards the Earth, they may cause severe space weather events, leading to the destruction of space and ground-based electronic devices \citep{Gosling1993,Webb1994}. 

White-light coronagraph data reveal that CMEs often exhibit a typical three-component structure: a leading front, a dark cavity and an embedded bright core \citep{Illing1986}. The CME leading front is usually explained as the plasma piled up at ahead of the erupting magnetic structure. The dark cavity and bright core are interpreted as magnetic flux rope (MFR) and cold prominence (filament) matter suspended in the magnetic dips of MFR, respectively. Nevertheless, white-light data are incapable of disclosing the early dynamics of CMEs, in particular, the formation of CMEs. Taking advantage of EUV multiple-wavelength imaging data, \cite{Zhang2012} and \cite{Cheng2013} found that the pre-eruptive configuration of CMEs first appears as a channel-like high temperature MFR structure. After being triggered, it quickly evolves towards the CME cavity and drives the formation of the CME leading front at the same time \citep{Cheng2014}. Moreover, the CMEs can also present a three-component structure in the EUV passbands, very similar to what is observed in white-light data \citep[e.g.,][]{Song2019b}.

\begin{figure*}[ht!]

{\centering\includegraphics[width=1.0\textwidth]{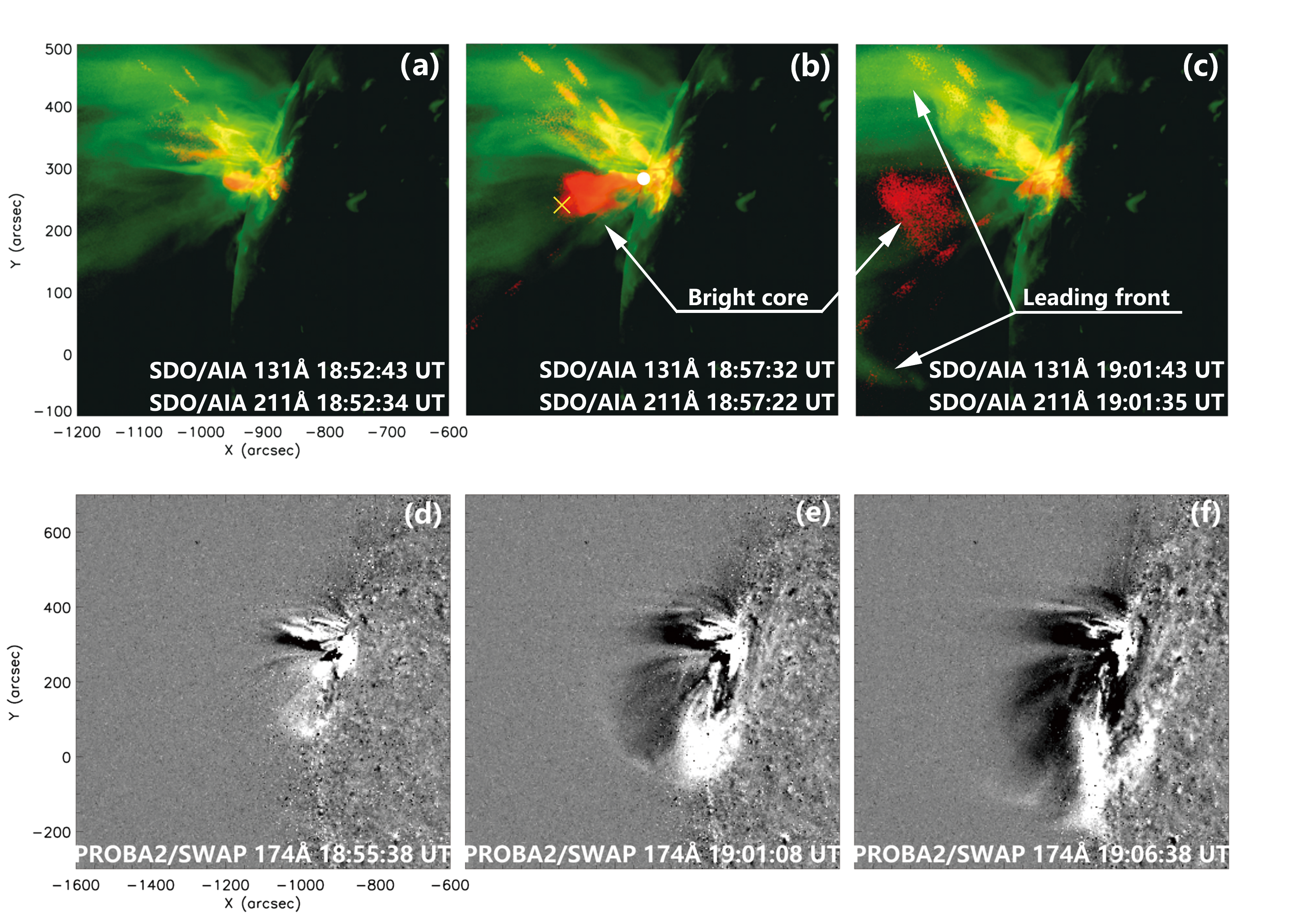}

}
\caption{(a)-(c) Composite images of the AIA 131 {\AA} (red) and AIA 211 {\AA} (green) showing the EUV leading front and bright core structure of the 2021 May 07 CME. In panel (b), the white point is the position of the source region and the yellow cross marks the top of the EUV blob. The leading front and bright core are pointed out in panels (b) and (c). The leading front is only partially visible in the AIA FOV. (d)-(f) SWAP 174 {\AA} base difference images showing the evolution of the CME leading front and bubble.\\
(Animations of AIA composite and SWAP images are available online.) \label{fig:1}}
\end{figure*}

The kinematic evolution of CMEs is closely coupled with the variation of the soft X-ray emission of associated flares \citep{cheng2020}. \cite{Zhang2001,Zhang2004} studied the temporal evolution of the velocity of flare-associated CMEs and found that it can be divided into three phases: a slow rise phase and an impulsive acceleration phase followed by a propagation phase of constant speed. The three evolution phases well correspond to the three phases of the soft X-ray flux of associated flares: a pre-flare phase, a rise phase and a decay phase, respectively \citep[also see][]{Zhang2001,Maricic2007}. Moreover, some studies revealed that the acceleration of CMEs coincides very well with the hard X-ray flux of associated flares \citep{Wang2003,Qiu2004,Jing2005,Temmer2008,Temmer2010}. These results strongly support the so-called standard flare model, in which CMEs and associated flares are believed to be two manifestations of the same physical process, i.e., reconnection-coupled disruption of the coronal magnetic field \citep{Lin2000,Priest2002}.

Apart from radial propagation, CMEs also present a lateral expansion, which could play a more important role in the CME formation. In the outer corona, it is found that the expansion of CMEs tends to be self-similar, i.e., expanding radially and laterally in the same rate \citep{Schwenn2005}. In the inner corona, however, CMEs will undergo strong lateral overexpansion, in which the CME width grows faster than that of the height of the CME centroid \citep{Patsourakos2010a,Patsourakos2010b,Cheng2013}. Such a strong overexpansion is even able to excite a CME shock wave at ahead of the CME leading front if the velocity exceeds the local Alfv$\acute{e}$n speed \citep{Veronig2008,Temmer2009,Patsourakos2010a,Veronig2010,Cheng2012}.

With appropriate conditions, the CME-driven shock wave can accelerate local electrons, which then give rise to type II solar radio bursts \citep{Krucker1999,Klassen2002,Simnett2002}. In radio dynamical spectrographs, type II bursts appear as a narrow band that drifts toward lower frequency, showing the evolution of the CME shock toward the region of lower density \citep{Wild1950}. On the other hand, type III radio bursts, presenting an extremely fast frequency drifting rate, appear in radio spectrographs more frequently. They are mostly argued to be caused by energetic electrons that are accelerated by associated flares and then propagate along open magnetic field lines. Some type III bursts even exhibit reversed drifting features, supposed to be caused by electrons moving toward the lower atmosphere \citep{van2008}.

Although many progresses have been achieved on the study of the CME early evolution, how the different structures of CMEs are formed is still far from fully understood. One of the main restrictions is the lack of observations that are capable of tracking CMEs continuously and routinely from the solar limb to 3 R$_{\sun}$. In this paper, we investigate a well-observed limb CME that took place on 2021 May 07 and was observed continuously from 0 to 3 R$_{\sun}$ by a combination of EUV and white-light images. In particular, we pay more attention to the CME overexpansion, as well as induced various radio bursts. In Section \ref{sec:instu}, we introduce instruments. In Section \ref{sec:obser}, we present the methods and main results, which are followed by a summary and discussions in Section \ref{sec:dis}.

\begin{figure*}[ht!]

{
\includegraphics[width=1.0\textwidth]{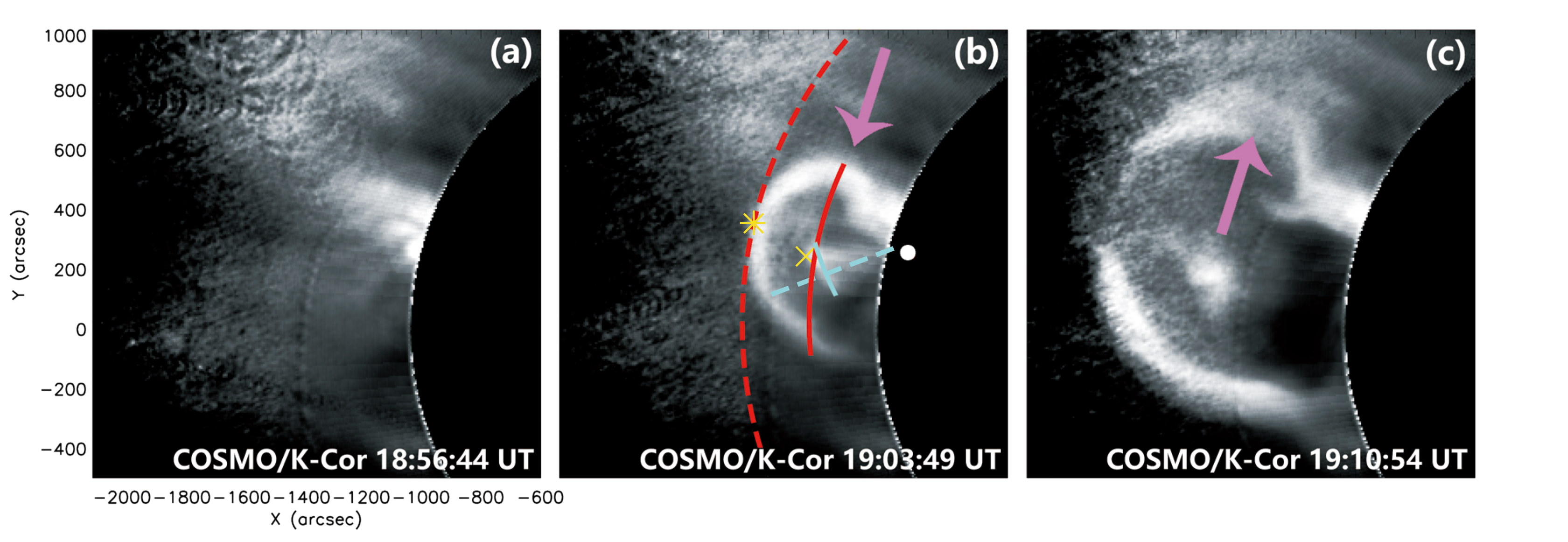}

}
\caption{K-Cor images of 2021 May 07 CME. The cyan dashed line shows the eruption direction of the bright core. The white point and the yellow cross in panel (b) have the same meaning as that in Figure \ref{fig:1}(b). The cyan solid line indicates the width of the bright core. The dashed line in red represents part of circle with its center at the solar center, which is tangent to the CME leading front, while the yellow asterisk marks its position. The solid line in red labels the width of the CME bubble. Purple arrows mark the positions of the interaction between the CME front and streamer. \\
(An animation of this figure is available.)
\label{fig:2}}
\end{figure*}

\section{Instuments} \label{sec:instu}
The Atmospheric Imaging Assembly \citep[AIA;][]{Lemen2012} on board Solar Dynamics Observatory (SDO) provides EUV images of the solar atmosphere with the field of view (FOV) out to 1.3 R$_{\sun}$ at seven EUV passbands (94, 131, 171, 193, 211, 304, 335 \AA), covering the temperature from 0.05 to 11 MK. The corresponding temporal candence and spatial resolution are 12 s and 1.5 arcsecs, respectively. The K-Coronagraph (K-Cor) located at Mauna Loa Solar Observatory (MLSO) observes the coronal polarization brightness in the white-light band with a FOV of 1.05--3 R$_{\sun}$.The spatial resolution is 11 arcsecs and the temporal candence is 15 s. In this paper, we mainly use the data from MLSO/K-Cor and SDO/AIA data at 131 \AA\ and 211 \AA\ passbands with the temporal candence of 60 s and 36 s, respectively. The K-Cor data we used have been well calibrated. The Sun Watcher using Active Pixel System detector and Image Processing \citep[SWAP;][]{Seaton2013} on board PRoject for Onboard Autonomy 2 (PROBA2) are also used as a supplement. The X-Ray Sensor \citep[XRS;][]{Chamberlin2009} on board Geostationary Operational Environmental Satellite-R (GOES-R) provides the soft X-ray 1--8 \AA\ flux of associated flare and the Spectrometer Telescope for Imaging X-rays \citep[STIX;][]{Krucker2020} on board Solar Orbiter (SO) observes the hard X-ray flux at the energy band from 4 to 150 keV. In addition, the combination of solar radio dynamic spectrum from 20 kHz to 100 MHz is obtained from Wind/Waves \citep{Bougeret1995} and e-CALLISTO \citep{Benz2009} network spectrometer at BIR station. 

\section{Observation} \label{sec:obser}

\subsection{Event overview} \label{subsec:ev}

On 2021 May 7, a limb CME taking place in Active Region (AR) NOAA 12822 was observed simultaneously by SDO, PROBA2 and MLSO. It originated from the eruption of a high-temperature blob, which was also named as CME core in the EUV passbands \citep{Song2019a}, as seen at the AIA 131 \AA\ passband (the red blob in Figure \ref{fig:1}a-\ref{fig:1}c). After its appearance ($\sim$18:50 UT), the blob started to grow up. At 18:57 UT, it seemed to rotate in a counterclockwise direction, presenting a channel-like feature, very similar to the hot-channel MFR as detected for face on events \citep[e.g.,][]{Zhang2012,Cheng2013,Cheng2014}. In the AIA 211 \AA\ and SWAP 174 \AA\ passbands, the erupting bubble caused the expansion of the overlying fields and the accumulation of plasma that quickly formed a bright leading front followed by a dark cavity (Figure \ref{fig:1}d-\ref{fig:1}e). The front and core structures of the CME as seen in the EUV passbands (Figure \ref{fig:1}b and \ref{fig:1}c) are very similar to what is usually observed in the white-light band \citep[e.g., Figure 3 and Figure 11 in][]{Veronig2018}.

\begin{figure*}[ht!]

{\centering\includegraphics[width=1.0\textwidth]{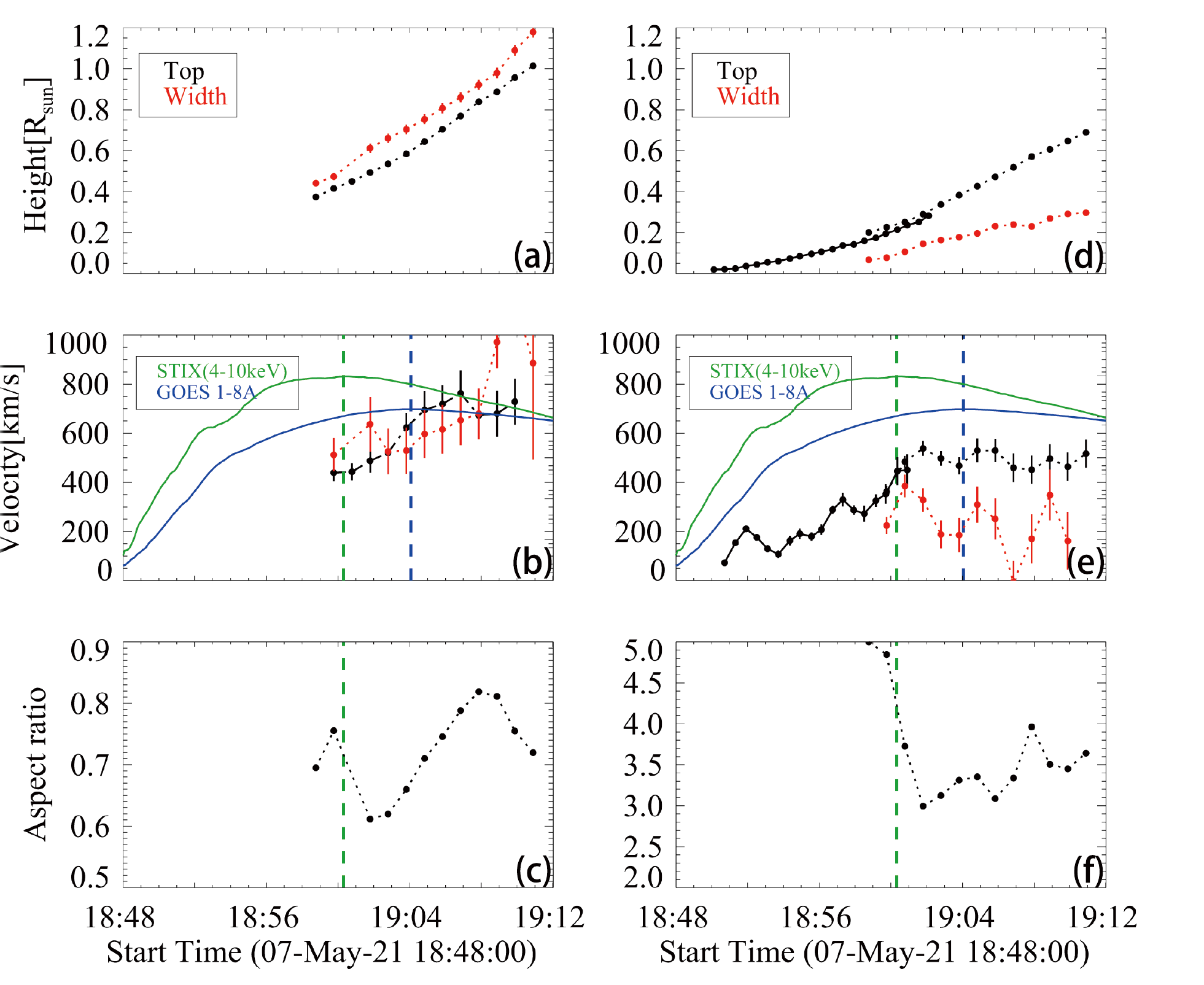}

}
\caption{(a) Temporal evolution of the height and width of the CME bubble. (b) Temporal evolution of the radial and expansion velocities with the vertical bars denoting their uncertainties. The STIX 4-10 keV and GOES 1-8 {\AA} flux are also plotted in green and blue, respectively, with the two vertical dashed lines marking their peaks. (c) The aspect ratio of the CME bubble. (d)--(f) are similar to (a)--(c) but for the CME bright core. The black dots connected by solid and dashed lines are the measurements from EUV blob and white light bright core, respectively. \label{fig:3}}
\end{figure*}

\begin{figure*}[ht!]

{\centering\includegraphics[width=0.8\textwidth]{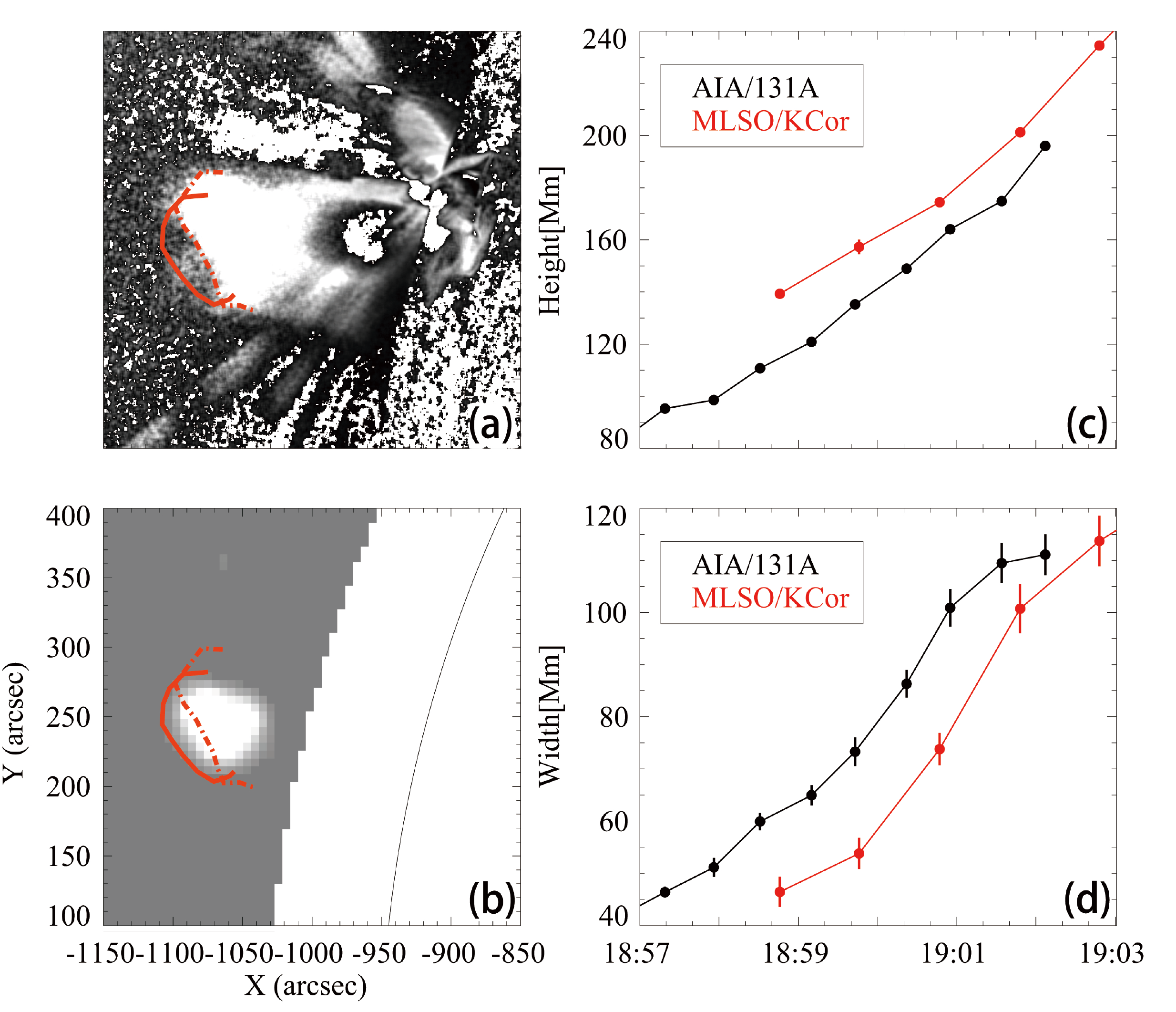}

}
\caption{(a) The AIA 131 {\AA} base difference image (at 18:59:43 UT) showing the EUV blob. The red dashed line delineates the top boundary of the EUV blob and the red solid line represents that of the white-light bright core. (b) The K-Cor base difference image (at 18:59:46 UT) showing the white-light bright core. The FOV of panel (a) is the same as that of panel (b). (c) Height-time profile of the CME EUV core and white-light core. (d) similar to (c) but for the width-time profile. \label{fig:4}}
\end{figure*}

Figure \ref{fig:2} shows the early evolution of the white-light CME captured by MLSO/K-Cor. The bright core was first observed at 18:56 UT. Similar to the erupting EUV high-temperature blob, it also propagated along a significantly non-radial direction before $\sim$19:11 UT. Nevertheless, we did not detect the rotation of the bright core as detected for the EUV blob. The CME leading front first appeared at $\sim$18:59 UT and propagated almost radially. Its top boundary can be identified clearly before $\sim$19:11 UT. One can find that the bright core appeared earlier than the leading front in the K-Cor FOV. the main reason is that at the early phase of the eruption, no enough plasma was accumulated at the CME front, which was thus invisible. Moreover, we find that the CME front was asymmetrical: its northern part seemed to be strongly compressed by a nearby streamer. Their interaction even gave rise to complicated radio bursts, which will be discussed in Section \ref{subsec:rad}. Unfortunately, the shock wave for the current event may be too weak to be clearly visible at the EUV and white-light bands

\subsection{Kinematics of the CME bright core and dark cavity} \label{subsec:ear}

To quantify the expansion and propagation of the different structures of the CME, we first define the height of the CME ($H_{f}$) as the radial distance from the highest point of the bright front (yellow asterisk in Figure \ref{fig:2}) to the solar limb that is independent of the CME position angle, the CME width ($D_{f}$) as the maximal arc length of the CME bubble relative to the heliocenter (red solid line in Figure \ref{fig:2}). 
Since the CME bright core propagated non-radially at the early stage, the definitions of its height and width are different from that for the CME bubble. The height ($H_{c}$) is the distance from the top of the CME bright core (yellow cross in Figures \ref{fig:1} and \ref{fig:2}) to the source region (white point in Figures \ref{fig:1} and \ref{fig:2}). The width ($D_{c}$) is the arc length (cyan solid line in Figures  \ref{fig:2}) in the direction perpendicular to its propagation direction (cyan dashed line in \ref{fig:2}). All the above parameters are measured five times manually.

Before doing measurements, we process the AIA images by the IDL routine $aia\_rfilter.pro$ in SSW package to enhance the coronal structures above the limb, to say, by summing a number of EUV images acquired at consecutive times to increase the signal-to-noise ratio. The off-limb corona is divided into concentric rings. The corona structures in different rings are scaled based on parameters including the ring radius, average brightness and intensity relative to the neighboring rings. For white-light coronagraph K-Cor data, the Normalized Radially Graded Filter \citep[NRGF;][]{Morgan2006,Morgan2012} is used, which scales the corona structure by the mean brightness and its standard deviation at different heights. Note that, since the CME leading front in the EUV images is not sharp enough, only the K-Cor data are used for quantifying the front parameters. 

The MFR at the AIA 131 \AA\ passband is found to be partially face-on, which makes it difficult to justify the cross section of the MFR; thus the core width is also derived from K-Cor data. For each parameter, we repeat the measurement five times and obtain the average value. The standard deviation is considered as the corresponding uncertainty. Then, through calculating the first-order numerical differential of height-time and width-time data by the IDL routine $deriv.pro$, we derive the eruption and expansion velocities of the CME bubble, named as $V_{Hf}$ and $V_{Df}$, respectively. For the CME bright core, the eruption and expansion velocities ($V_{Hc}$ and $V_{Dc}$) are also derived similarly. Moreover, we define the aspect ratio to identify the overexpansion, which is the height of the CME (core) centroid divided by its half-width. For simplicity, we use $H_{f}-0.5D_{f}$ and $H_{c}-0.5D_{c}$ to denote the height of CME and core centroids, respectively. The aspect ratios are then derived by $r_{f}=(2H_{f}-D_{f})/D_{f}$ and $r_{c}=(2H_{c}-D_{c})/D_{c}$ for the CME bubble and the core, respectively. To further reveal the temporal relationship between the CME overexpansion and the flare energy release, besides the  GOES 1-8\AA~ soft X-ray flux, the lower energy part (4-10 keV) of the STIX hard X-ray flux is also used, which is mainly produced by the bremsstrahlung of thermal electrons.

Figure \ref{fig:3} shows the early temporal evolution of the CME bubble (panels a--c) and bright core (panels d--f) in the radial and lateral directions.
The width and height of the CME bubble are plotted in Figure \ref{fig:3}(a) as a function of time, from which one can find the width of the CME bubble is larger than its height during the observation period. Figure \ref{fig:3}(b) shows the evolutions of the radial and lateral velocities overplotted by the STIX 4-10 keV and GOES 1-8\AA~ flux curves. Both of them increase continuously before 19:08 UT. After that, the radial velocity tends to be a constant of 600-700 km s$^{-1}$, while the lateral velocity still quickly increases from $\sim$600 km s$^{-1}$ up to $\sim$1000 km s$^{-1}$ within 2 minutes. In Figure \ref{fig:3}(c), the aspect ratio clearly shows two decrease phases. The former one takes place near the flare peak time, i.e., the impulsive energy release of the flare revealed by STIX 4-10 keV flux, in a good agreement with the finding of \cite{Cheng2013}. However, after an increase phase, the aspect ratio decreases again (after 19:08 UT) even though the flare has already evolved into the decay phase. The results show that the height increment of the CME leading front in the early phase is mostly from its expansion.

The bright core of the CME appearing as the EUV high-temperature blob was tracked continuously to 0.3 R$_{\sun}$. After that, it appeared at the white-light band and was captured by the K-Cor from 0.2 to 0.7 R$_{\sun}$. The temporal evolutions of its height and width are presented in Figure \ref{fig:3}(d) and that of the corresponding velocities in Figure \ref{fig:3}(e). One can find that, the eruption velocity of the bright core gradually increases to over 500 km s$^{-1}$ near the flare peak (19:02 UT) and then maintains a constant in the flare decay phase. The expansion velocity of the bright core also reaches its maximum near the flare peak and subsequently starts to decrease, which obviously differs from that for the CME bubble.
The aspect ratio of the bright core rapidly decreases in the time period of 18:58 UT-19:01 UT. Similar to the CME bubble, the minimal aspect ratio is found to roughly correspond to the peak of the STIX 4-10 keV flux. It indicates that the CME bright core also experienced a strong lateral expansion but only during the main energy release process of the flare. Note that, the aspect ratios of the CME bubble and bright core are greatly different, which is mainly due to the fact that their widths are obviously distinct but the centroid heights are similar.

\begin{figure*}[ht!]

{\centering\includegraphics[width=1.0\textwidth]{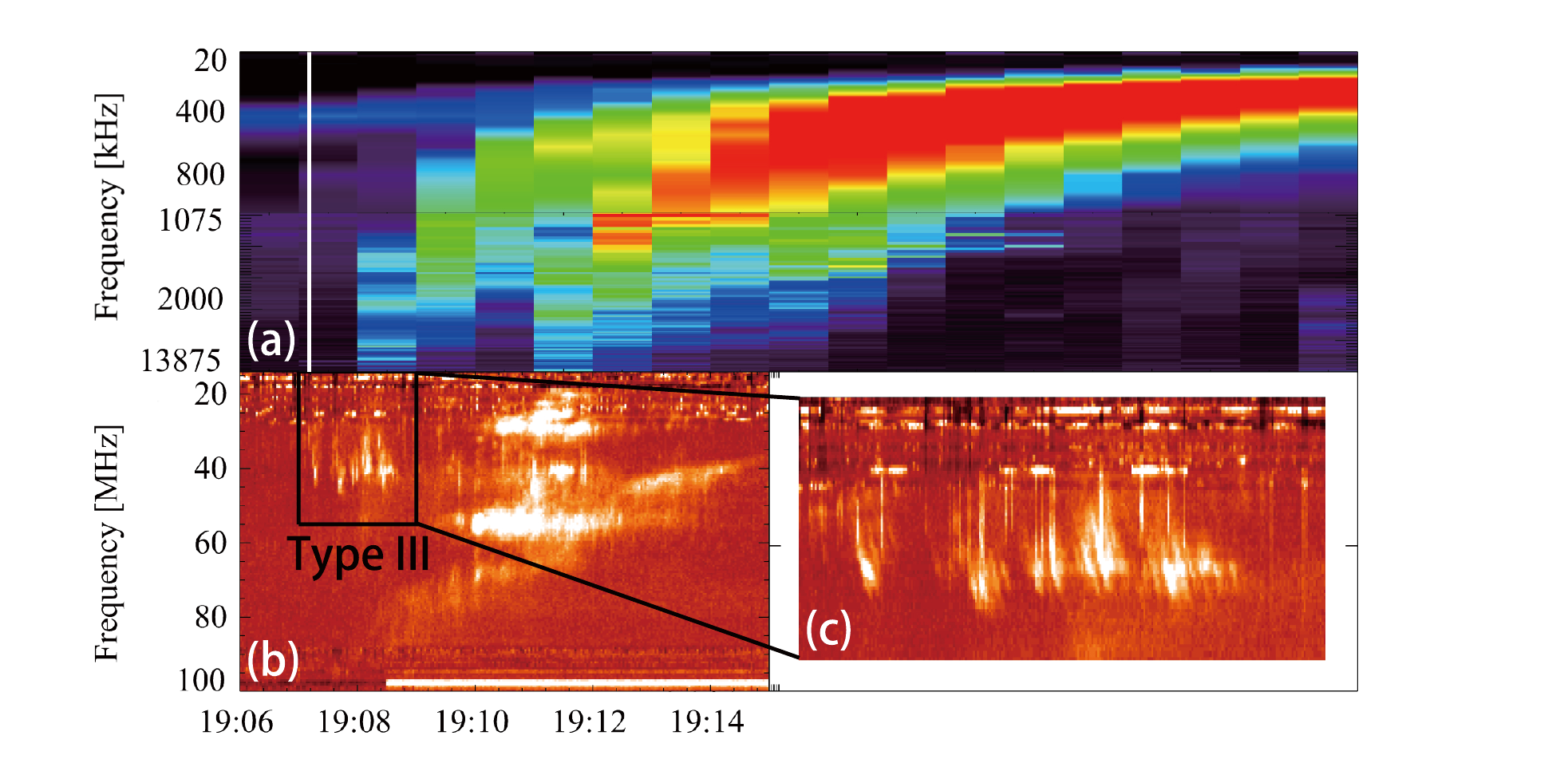}

}
\caption{(a) Wind/Waves radio dynamic spectrograph from 20 to 13875 kHz, the white solid line marks the start time of the type III burst. (b) The dynamic spectrograph in the frequency range from 14 to 100 MHz from BIR station. (c) Zoom in of a group of type III bursts.  \label{fig:5}}
\end{figure*}

\subsection{EUV MFR and white-light bright core} \label{subsec:com}
The overlapped FOV of the AIA and K-Cor allows us to explore how the different EUV structures of the CME evolve toward the white-light structures. Here, we mainly focus on the evolution of the erupting high-temperature blob. As shown in Figure \ref{fig:4}, we find that the EUV blob seems to be wider but lower than the white-light bright core at the same moment (Figure \ref{fig:4}(c) and \ref{fig:4}(d)). Here, we determine the boundary of the bright core by performing the threshold segmentation on the original data (Figure \ref{fig:4}b). We set the threshold to be 40\% of the maximal brightness. In order to check the influence of the threshold, we also change the threshold within 20\% and find that it does not affect our conclusion qualitatively.

Figure \ref{fig:4}c and \ref{fig:4}d show the temporal evolution of the height and width of the EUV blob and that for the white-light bright core, respectively, during the time period of 18:57 UT--19:02 UT. The discrepancy in height between the two structures is over 10 Mm, similar to the discrepancy in width. Note that, we here use the same method to measure the height and width of the EUV blob and white-light core as that for Figure \ref{fig:3}.
Such a discrepancy mainly arises from the fact that the visibility of the EUV MFR structure is not only related to the density but also the temperature, while that of the white-light structure only relies on the density.

Furthermore, for this particular event, we do not observe an erupting filament, and thus the CME white-light bright core is likely contributed by the erupting flux rope as argued by \citet{Song2019a}.
However, during the flare, magnetic reconnection proceeds continuously, forming the poloidal flux that is added to and envelopes the erupting MFR. Meanwhile, due to the reconnection heating, the envelope layer of the flux rope should be much hotter than its inner part \citep[e.g.,][]{Cheng2018}. 
On the other hand, the density of the MFR drastically decreases as it erupts outward because of its overexpansion.
We speculate that the entire MFR could be larger than both the observed EUV blob and white-light bright core. The discrepancy between the morphologies shown in EUV and white-light is mainly caused by their different responses to the density and temperature structures within the MFR. In addition, there are many other factors that also influence the visibility of CME different structures, such as the projection effect and the deflection during the propagation.

\subsection{Radio bursts caused by CME overexpansion}
\label{subsec:rad}
\begin{figure*}[ht!]

{\centering\includegraphics[width=1.0\textwidth]{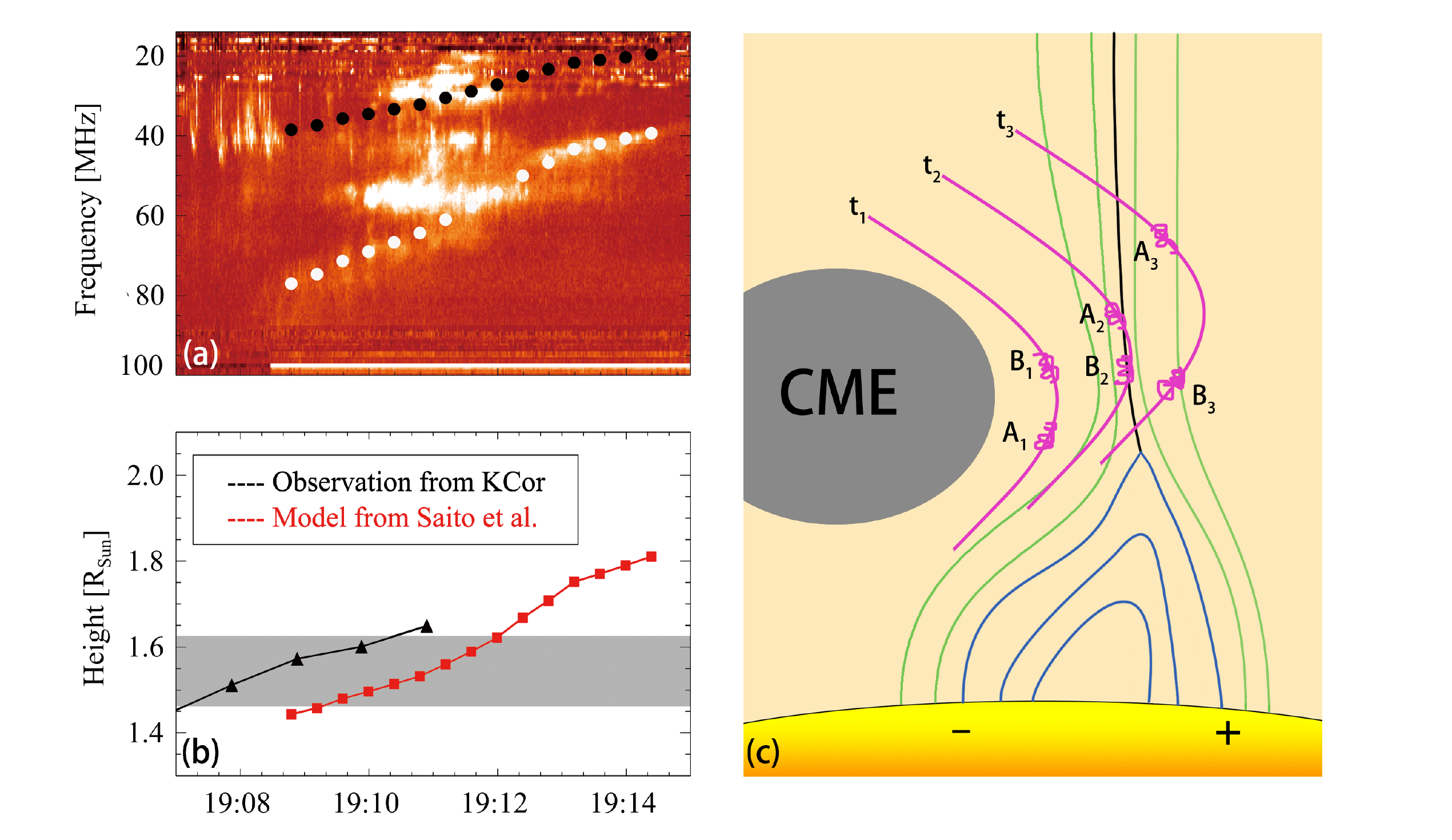}

}
\caption{(a) The radio dynamic spectrograph representing a drifting and stationary type II radio burst. The white (black) points mark the harmonic (fundamental) frequency of the drifting type II burst and are used for estimating the heights of the source region. (b) Height-time profiles of the drifting type II burst source region derived from the model of \cite{Saito1977} (red) and of the interaction region between the CME and streamer estimated directly using the K-Cor data (black). The shaded region represents the possible initial height of the type III bursts. (c) A sketch interpreting the interaction between the CME shock and streamer. The green and blue curves are the opened and closed streamer magnetic field lines, the purple curves represent the CME-driven shock front and the black curve is the streamer current sheet. The radio-emitting regions related to the type II bursts are marked as $A_{i}$ and $B_{i}$. \label{fig:6}}
\end{figure*}

As the CME expanded, it interacted with the streamer in the north and caused three kinds of radio bursts including a normal type II burst, a stationary type II burst and a group of type III bursts with both normal and reversed parts, whose dynamic spectrographs are shown in Figure \ref{fig:5}. In Figure \ref{fig:5}(b), one can clearly see that the normal type II burst consisted of one fundamental and one harmonic components, the latter of which started at 80 MHz at 19:08 UT and slowly drifted to 40 MHz at 19:15 UT with a drifting rate of 1 MHz/s (also see Figure \ref{fig:6}). Moreover, the stationary type II burst also included fundamental and harmonic components, which appeared at 27 MHz and 55 MHz, respectively, and lasted for about 5 minutes. In Figure \ref{fig:5}(c), the zoom-in dynamic spectrograph at the time period of 19:07 -19:09 UT clearly displays the reversed type III signals: a group of type III bursts started at about 30 MHz and then drifted toward the higher frequency of over 40 MHz. In Figure \ref{fig:5}(a), there are also signals towards $\sim$100 kHz that lasted for more than 20 minutes. Both the positive and negative drifting type III bursts form a "C" shape in the spectrum, thus named as "C-shaped" type III bursts thereafter.

To reveal the origin of these radio bursts, we study the interaction of the CME overexpansion with the northern streamer. Taking advantage of the density model, we first estimate the height of the source region of the drifting type II burst. The frequency is converted to the density based on the following formula: 
\begin{equation} \label{eq:1}
f_{p} = 8.98 \times 10^{-3} \sqrt{n_{e}}
\end{equation}
where, $f_{p}$ and $n_{e}$ represent the frequency (MHz) and the electron density (cm$^{-3}$), respectively. The plasma frequency is taken from the half of the first harmonic component, since it appears more clearly than the fundamental one and thus can be used in measurements with a higher precision.
The density model we use here is from \cite{Saito1977}, which takes the following form:
\begin{equation} \label{eq:2}
n_{e} = 1.36 \times 10^{6}\,r^{-2.14} + 1.68 \times 10^{8}\,r^{-6.13}
\end{equation}
where $r$ is in units of solar radii and $n_{e}$ in units of cm$^{-3}$. Solving the equations (\ref{eq:1}) and (\ref{eq:2}) by non-linear least squares method, we derive the height of the type II source region and its temporal evolution. The results are shown in Figure \ref{fig:6}b, in which we also overplot the heights of the CME flank (where we determine the CME width in Figure \ref{fig:2}) as the region of interaction with the streamer (the black triangles in Figure \ref{fig:6}b). One can see that the two heights are very close to each other,
indicating that the type II burst is most likely produced at the interaction regions where the CME compresses the streamer. 

With regard to the stationary type II burst, it may originate from a different source from the drifting type II burst. \citet{Aurass2004} ever observed a zero-drifting type II burst and argued that it was caused by the interaction between a termination shock driven by the reconnection outflows and the flare loops. However, obviously, the frequency of the zero-drifting type II burst under study is much lower than that observed by \citet{Aurass2004}. We speculate that this is also generated during the CME-driven shock crossing the streamer. In Figure \ref{fig:6}c, we interpret the possible generation process of two type II bursts. It is likely that two sources of type II bursts are formed during the interaction of the CME-driven shock and streamer with one (A) moving outward and the other (B) just crossing the streamer, which generate the drifting and stationary type II bursts, respectively. 

The type III radio burst is believed to be caused by energetic electrons that escape away from the Sun along the open field lines. However, except for the normal type III burst, we also observe a group of reversed type III bursts, which had a starting frequency of $\sim$30 MHz and then drifted toward the higher frequencies (Figure \ref{fig:5}c), 
showing that energetic electrons propagate downward. 
Based on the density model, the initial height of the reversed type III bursts is estimated to be in the range of 1.45-1.61 R$_{\sun}$ (the gray region in Figure \ref{fig:6}b), which is very close to the height where the CME interacts with the streamer. Moreover, the reversed type III bursts mostly appeared near the onset time of the drifting type II burst. Thus, it is most likely that the interaction between the CME and streamer also involves interchange reconnection, which releases energetic electrons trapped in the erupting CME. These electrons move upward and downward, giving rise to positive-drifting and negative-drifting type III bursts, respectively.

\section{Summary and Discussions} \label{sec:dis}
In this paper, we study the formation and early dynamics of a CME on 2021 May 07. The CME occurred near the solar limb and was captured by SDO/AIA and MLSO/K-Cor simultaneously, thus allowing us to investigate the continuous evolution of the CME different structures from 0 to 3 R$_{\sun}$ by minimizing projection effect. An interesting finding is that this CME bubble underwent two overexpansion processes. The first overexpansion process occurred near the flare peak, similar to the previous results \citep{Patsourakos2010a,Cheng2013,Balmaceda2022}, and the second one in the decay phase, when the magnetic reconnection has greatly weakened.

Based on the standard flare model \citep{Carmichael1964,Sturrock1966,Hirayama1974,Kopp1976}, \cite{Patsourakos2010a} proposed two effects that probably cause the overexpansion. First, the reconnection in the flare current sheet underneath the MFR continuously adds extra poloidal flux to the erupting MFR, thus causing the MFR overexpansion. Second, an MHD process (called inverse pinch effect in \cite{Kliem2014}) is also capable of causing the overexpansion. As the MFR rises, the current decreases because of the flux conservation between the MFR and the photosphere. As a result, the poloidal flux decreases and a larger volume is required to compensate for the weaker fields, acting as the lateral overexpansion of the MFR. Based on the relation of the CME overexpansion to the flare, for the current event, the first and second overexpansion is attributed to the reconnection-dominated effect and the ideal MHD inverse pinch effect, respectively. In \cite{Balmaceda2022}, the overexpansion of two bubble-like CMEs was also interpreted to be caused by the reconnection process based on its simultaneity with the inpulsive phase of associated flare. \cite{Zhuang2022} even found the CME overexpansion extending to the high corona over 10 R$_{\sun}$. However, different from our interpretation, they argued that the flare reconnection was the unique mechanism to drive the overexpansion continuously.

Thanks to the overlapped FOV of AIA and K-Cor, we compare the morphological discrepancy between the white-light bright core and EUV blob. The EUV blob is lower but wider than the bright core. In the context of standard flare model, a large quantity of heat flux is generated quickly by the flare reconnection and then added to the MFR, thus resulting in the MFR fast expansion \citep{Cheng2013,Veronig2018}. However, the temperature of the MFR could be nonuniform. The lower part of the MFR is found to be much hotter than the upper part \citep{Cheng2018,Ye2021}; thus probably only part of the MFR is visible at the EUV high-temperature passbands. It means that the real CME MFR is most likely larger than the EUV blob, in particular when the hot MFR has cooled down; while for the white-light bright core, it merely corresponds to a part of the MFR where the plasma is dense.

The interaction of the expanding CME with the nearby streamer produced multiple radio bursts including a typical and a stationary type II burst, as well as a "C-shaped" type III burst. We argue that the two type II bursts are generated at two sources that propagate in the different directions at the flank of the CME-driven shock. This is similar to the interpretation of \cite{Balmaceda2022} for the different normal type II radio burst that was also generated by the CME-overexpansion-driven shock. As shown in Figure \ref{fig:6}(c), the CME-driven shock may be strengthened during the interaction with the streamer, mainly due to the decrease of the local Alfv$\acute{e}$n velocity caused by the increase in density. The strengthened shock then accelerates electrons at the position A and B that propagate upward and laterally, thus resulting in a drifting and a stationary type II burst, respectively.
In fact, a stationary type II caused by the interaction between the CME and streamer is also suggested by \cite{Chrysaphi2020}, in which the streamer is believed to stop the moving CME in the early stage and the frequency of the radio source hence does not change with time. However, as the CME moves, the stationary type II burst starts to transit into the drifting type II burst. Obviously, such a situation is quite different from what observed here. In our case, the normal and stationary type II bursts occur simultaneously. They are thus most likely from the two different sources at the CME-driven shock front.

The interchange reconnection may take place at the beginning when the expanding CME interacts with the streamer. This is strongly supported by the appearance of a group of "C-shaped" type III bursts. In general, the formation of type III bursts needs two conditions, one of which is accelerated electrons and the other is open flux. Nevertheless, the energetic electrons are mostly restricted within the close flux of the erupting CME unless the reconnection takes place between the CME flux and the background field so as to release restricted electrons \citep{van2008,Masson2013,Kou2020}. The released electrons propagate upward and downward and then give rise to the "C-shaped" type III burst \citep{van2008,Hillaris2011,Zheng2017}.

\begin{acknowledgements}
This work is supported by NSFC grants 11722325, 11733003, 11790303, 12127901 and National Key R\&D Program of China under grant 2021YFA1600504. X.C. is also supported by Alexander von Humboldt foundation.
\end{acknowledgements}

\bibliographystyle{aa}
\bibliography{articles}
\end{document}